\documentclass[prb,preprint,amsmath,amssymb,showpacs,superscriptaddress]{revtex4}
\usepackage{graphicx}
\usepackage{dcolumn}
\usepackage{bm}
\usepackage{hyperref}%

\hypersetup{colorlinks=true,citecolor=blue,urlcolor=blue,linkcolor=blue}

\def\dc{$^{\circ}$}

\begin{document}


\title{Electrical transport properties of polar heterointerface between KTaO$_3$ and SrTiO$_3$}

\author{A. Kalabukhov}
  \email{alexei.kalaboukhov@mc2.chalmers.se}
  \affiliation{%
Department of Microtechnology and Nanoscience (MC2), Chalmers University of Technology\\
G\"oteborg, Sweden
}%
\author{R. Gunnarsson}
    \affiliation{%
Department of Microtechnology and Nanoscience (MC2), Chalmers University of Technology\\
G\"oteborg, Sweden
}%

\author{T. Claeson}
  \affiliation{%
Department of Microtechnology and Nanoscience (MC2), Chalmers University of Technology\\
G\"oteborg, Sweden
}%
\author{D. Winkler}
\affiliation{%
Department of Microtechnology and Nanoscience (MC2), Chalmers University of Technology\\
G\"oteborg, Sweden
}%

\date{\today}

\begin{abstract}
Electrical transport of a polar heterointerface between two
insulating perovskites, KTaO$_3$ and SrTiO$_3$, is studied. It is
formed between a thin KTaO$_3$ film deposited on a top of
TiO$_2$-terminated (100) SrTiO$_3$ substrate. The resulting
(KO)$^{1-}$(TiO$_2$)$^0$ heterointerface is expected to be
hole-doped according to formal valences of K (1+) and Ti (4+). We
observed electrical conductivity and mobility in the
KTaO$_3$/SrTiO$_3$ similar to values measured earlier in
electron-doped LaAlO$_3$/SrTiO$_3$ heterointerfaces. However, the
sign of the charge carriers in KTaO$_3$/SrTiO$_3$ obtained from
the Hall measurements is negative. The result is an important clue
to the true origin of the doping at perovskite oxide
hetero-interfaces.
\end{abstract}

\maketitle

The mechanism of doping in hetero-interfaces between two
insulating perovskite oxides has been intensively discussed since
the observation of large electrical conductivity in the
hetero-structure between LaAlO$_3$ (LAO) and SrTiO$_3$ (STO).
\cite{Ohtomo2004,Huijben2006,Thiel2006,Kalabukhov2007,Herranz2006,Siemons2006}
It was argued that when a thin LAO film is coherently grown on the
TiO$_2$-terminated STO substrate, the resulting interface
(LaO)$^+$(TiO$_2$)$^0$ is expected to be polar, provided that the
bulk formal valences of Ti and La are maintained at the interface
and that the structure of the interface is atomically perfect. The
polar structure at the interface results in an infinitely growing
electrostatic potential in the (001) direction when the thickness
of LAO film is increasing. In order to compensate for the charge
discontinuity at the interface, half of an electron per square
unit cell may be released leading to conductivity at the
interface. \cite{Nakagawa2006} The possibility of this doping
mechanism was supported by theoretical works.
\cite{Pentcheva2006,Park2006} However, there are other possible
doping mechanisms in perovskite oxides. It is well known that
electrical properties of STO can be changed from insulating to
metallic ones by a small reduction of oxygen from its
stoichiometric composition. \cite{Tufte1967,Koonce1967} The
possibility that the electrical property at the LAO/STO
heterointerface is not due to the oxygen vacancies was presumably
ruled out by keeping the STO substrate at deposition conditions
which does not result in bulk conductivity due to oxygen
vacancies. However, it is still possible that the deposition of
the LAO film itself can reduce oxygen from a shallow layer at the
STO substrate, as argued by Siemons et al. \cite{Siemons2006} We
have previously suggested that oxygen vacancies play an important
role in the electrical properties of the LAO/STO heterointerface.
\cite{Kalabukhov2007} However, the true microscopic origin of the
conductivity at the interface between LAO and STO could not be
understood.

In this work we treat another polar interface between two
insulating perovskite oxides, KTaO$_3$ (KTO) and STO. KTO is a
well known material with a cubic structure and lattice parameter
of 3.99 \AA (compare with 3.905 \AA in STO). It is incipient
ferroelectric at room temperature with a dielectric constant of
about 260. \cite{Bozinis1976, Bae2004} Tantalum has a formal
valence of 5+, and potassium 1+ in KTO. The KTO film should grow
as a sequence of layers on a single TiO$_2$-terminated STO
substrate in the (001) direction and the resulting interface
should have the structure of (KO)$^-$(TiO$_2$)$^0$. This means
that half a hole per square unit cell should be released. This is
opposite to the (LaO)$^+$(TiO$_2$)$^0$ heterointerface,  where
half an electron per unit cell is transferred to the interface.

We have grown thin KTO films on STO substrates and found that the
KTO/STO interface is indeed conducting with electrical properties
very similar to the LAO/STO interface. However, the charge of
electrical carriers deduced from Hall effect measurements is
negative. We discuss possible reasons for this interesting result
in view of interface structure and possible doping mechanisms.

\begin{figure}
\includegraphics[keepaspectratio,width=2.5 in]{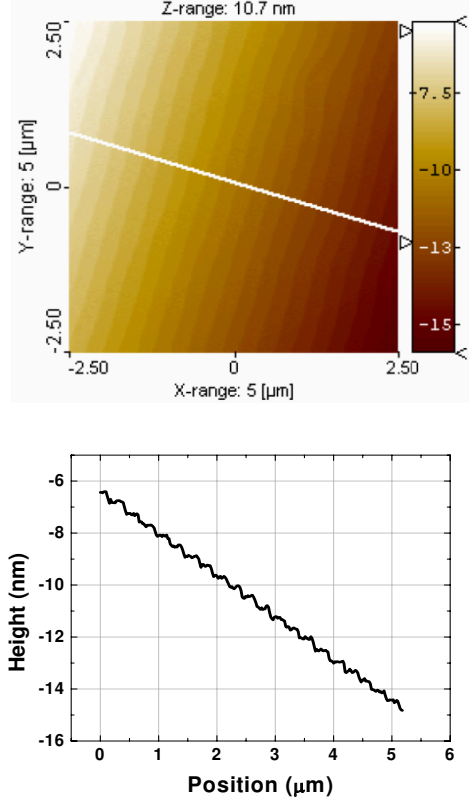}
\caption{\label{Fig1} (Color online) Atomic force microscope image
(top) and cross section (bottom) of the surface of the 6 nm thick
KTaO$_3$ film grown on a TiO$_2$-terminated (100) SrTiO$_3$
substrate. Unit cell steps are seen about every 250 nm along the
surface.}
\end{figure}

KTO films were prepared by pulsed laser deposition with in-situ
reflection high-energy electron diffraction (RHEED) used to
monitor film growth during deposition. The growth conditions were
similar to what we used previously to fabricate LAO/STO
hetero-interfaces: \cite{Kalabukhov2007} deposition temperature
T$_D$ = 750 \dc C, oxygen pressure p$_{O2}$ = $10^{-4}$ mbar,
laser energy density J = 1.5 J/cm$^2$. RHEED oscillations could be
observed during the initial part of the film growth. However the
intensity decreased rapidly and after 3 unit cells it was too low
to observe oscillations. The deposition rate estimated from the
first RHEED oscillations was 1 unit cell per 10 pulses. The
thickness of the KTO films was 13 u.c. layers (approx. 6 nm).
Atomic force microscopy (AFM) showed very smooth step-like surface
of the KTO film, see Fig.\ref{Fig1}.

Electrical measurements were made in a four point van der Pauw
configuration~\cite{Pauw1958} in the temperature range
2~K~--~300~K and in magnetic field up to 5~T. First we proved that
the KTO film itself is not conducting by using "soft" contacts:
silver wires glued on the film surface using silver epoxy. In
order to reach the interface, we used Ti/Au contact pads
fabricated by sputtering through metal mask. The resistance
between Ti/Au contacts and contacts glued by silver epoxy was
above 10 M$\Omega$, indicating an absence of pinholes in the KTO
film.

\begin{figure*}
\includegraphics[keepaspectratio,width=6 in]{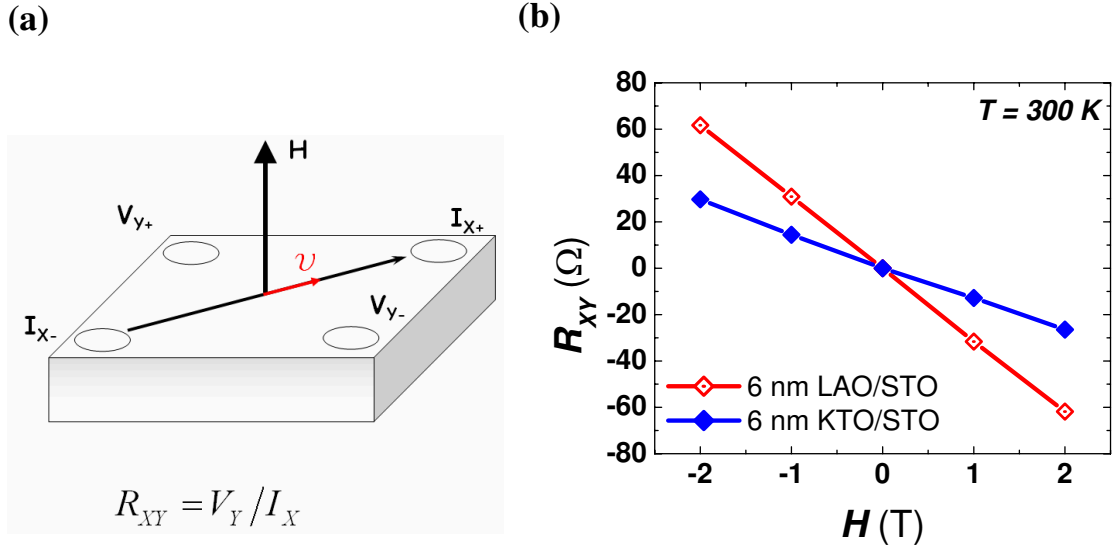}
\caption{\label{Fig2} (Color online) (a) Experimental
configuration for determination of Hall coefficient; (b) Hall
resistance R$_{XY}$ for KTO/STO and LAO/STO heterostructures
measured at room temperature and the same experimental
configuration.}
\end{figure*}

The electrical properties of KTO films may be compared to those of
15 u.c. thick LAO films on TiO$_2$-terminated STO substrates
prepared in the same conditions. Both hetero-structures show
metallic conductivity with relatively high mobilities and charge
carrier concentrations. Fig.\ref{Fig2} shows Hall resistances
measured at room temperature under the same experimental
configuration (i.e. magnetic field and current direction, see
Fig.\ref{Fig2}a). Both KTO/STO and LAO/STO heterointerfaces had
the same sign of Hall coefficient.  The sign of the charge
carriers is negative according to the magnetic field and bias
current directions.

\begin{figure*}
\includegraphics[keepaspectratio,width=6.5 in]{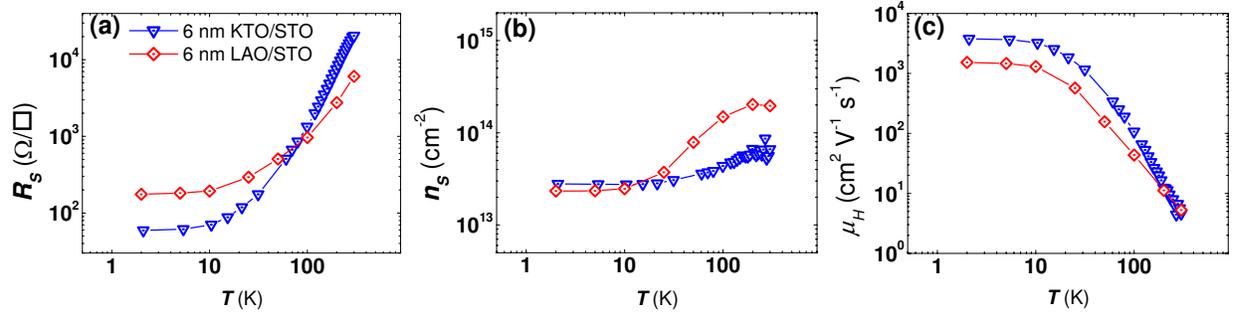}
\caption{\label{Fig3} (Color online) Temperature dependence of
sheet resistivity $R_{S}$ (a), charge carriers density $n_{S}$ (b)
and Hall mobility $\mu_{H}$ for LAO/STO and KTO/STO
heterointerfaces prepared at $10^{-4}$ mbar oxygen pressure.}
\end{figure*}

The values of the sheet resistance $R_{S}$, the Hall mobility
$\mu_{H}$ and the charge carrier density $n_{S}$ of the KTO/STO
heterointerfaces are very similar to those of LAO/STO, see
Fig.\ref{Fig3}. We measured three KTO films prepared in similar
deposition conditions and they all showed similar transport
properties.

It is known that potassium deficiency is a significant problem in
growth of KTO films  due to the high vapor pressure of potassium
at high temperatures. \cite{Bae2004} If this were the case here,
the actual heterointerface between KTO/STO may have different
reconstruction from the one described above. This possibility
needs to be ruled out in a future determination of the
microstructure of the hetero-structure by electron microscopy.

Independent of the KTO/STO heterointerface microstructure being
perfect or not, it is quite remarkable that the electrical
properties are very similar to those of LAO/STO heterointerface.
This could suggests that there is a common doping mechanism where
the type and concentration of charge carriers do not directly
depend on the sign of the polar interface deduced from the formal
bulk valences. We have previously argued that the high
conductivity, mobility, and charge carrier density found in
hetero-junctions of LAO/STO prepared at low oxygen pressure mainly
are due to oxygen vacancies residing in STO close to the
interface. That conclusion is further strengthened by the present
findings.

The work was supported by the Swedish KAW, SSF, and VR
foundations, the EU NANOXIDE, and ESF THIOX programs.


\begin{thebibliography}{1}
{
\bibitem{Ohtomo2004} A. Ohtomo, and H.Y. Hwang,
Nature
 \textbf{427}, 423 (2004);
 \textbf{441}, 120 (2006).

\bibitem{Huijben2006} M. Huijben, G. Rijnders,D. H. A. Blank, S. Bals, S. van Aert, J. Verbeeck,
G. van Tendeloo, A. Brinkman and H. Hilgenkamp, Nature Materials
 \textbf{5}, 556 (2006).

\bibitem{Thiel2006}S. Thiel, G. Hammerl, A. Schmehl, C. W.Schneider and
J. Mannhart,
Science
 \textbf{313}, 1935 (2006).

\bibitem{Kalabukhov2007} A. Kalabukhov, R. Gunnarsson, J. B\"{o}rjesson, E. Olsson, T.
Claeson and D. Winkler,
 Phys.\ Rev.\ B
 \textbf{75}, 121404(R) (2007).

\bibitem{Herranz2006} G. Herranz, M. Basletic, M. Bibes, R. Ranchal, A. Hamzic,
E. Tafra, K. Bouzehouane, E. Jacquet, J. P. Contour, A. Barthelemy
and A. Fert, Phys.\ Rev.\ B
 \textbf{73}, 064403 (2006).

\bibitem{Siemons2006} W. Siemons, G. Koster, H. Yamamoto, W. A. Harrison, G. Lukovsky, T. H. Geballe, D. H. A. Blank and M. R.
Beasley,
  cond-mat/0612223 (2006).

\bibitem{Nakagawa2006} N. Nakagawa, H. Y. Hwang and D. A. Muller,
Nature Mat.,
 \textbf{5}, 204 (2006).

\bibitem{Pentcheva2006} R. Pentcheva and W. E. Pickett,
Phys. Rev. B \textbf{74}, 035112 (2006)

\bibitem{Park2006} M. S. Park, S. H. Rhim and A. J. Freeman,
Phys. Rev. B \textbf{74}, 205416 (2006)

\bibitem{Tufte1967} O. N. Tufte and P. W. Chapman,
Phys.\ Rev.
 \textbf{155}, 796 (1967).

\bibitem{Koonce1967} C. S. Koonce, M. L. Cohen, J. F. Schooley, W. R. Hosler and E. R. Pfeiffer,
Phys.\ Rev.
 \textbf{163}, 380 (1967).

\bibitem{Bozinis1976} D. Z. Bozinis, J. P. Hurrel,
Phys. Rev. B
 \textbf{13}, 3109 (1976).

\bibitem{Bae2004} H.-J. Bae, J. Sigman, S.-J. Park, Y.-H. Heo, L. A.
Boatner and D. P. Norton,
 Solid State Electronics
 \textbf{48}, 51 (2004).

\bibitem{Pauw1958} J. L. van der Pauw
Philips Res.\ Rep.
 \textbf{13}, 1 (1958).
 }

\end{thebibliography}
\end{document}